\documentclass[a4paper]{jpconf}
\usepackage{graphicx}
\usepackage{color}
\usepackage[tight,TABTOPCAP]{subfigure}
\usepackage{latexsym}
\usepackage{amsmath}

\begin{document}

\title{$Z^\prime tc$ coupling from $D^{0}-\overline{D^{0}}$ mixing}

\author{J. I. Aranda$^{(a)}$, F. Ram\'\i rez-Zavaleta$^{(a)}$, J. J. Toscano$^{(b)}$ and E. S. Tututi$^{(a)}$}

\address{$^{(a)}$Facultad de Ciencias F\'\i sico Matem\' aticas,
Universidad Michoacana de San Nicol\'as de
Hidalgo, Avenida Francisco J. M\'ujica S/N, 58060, Morelia, Michoac\'an, M\' exico. \\
$^{(b)}$Facultad de Ciencias F\'{\i}sico Matem\'aticas,
Benem\'erita Universidad Aut\'onoma de Puebla, Apartado Postal
1152, Puebla, Puebla, M\'exico.}

\ead{jarandas@umich.mx}

\begin{abstract}
We bound the $Z^\prime tc$ coupling using the $D^{0}-\overline{D^{0}}$ meson mixing system. We obtained such coupling which is less than $5.75\times 10^{-2}$. We have studied the $Z^\prime$ boson resonance considering single top production in the $e^+e^-\to Z^\prime\to tc$ process. We obtained the number of events which is expected to be less than $10^7$ at the International Linear Collider scenario. We get a branching ratio of the order of $10^{-2}$ for the $Z^\prime\to tc$ decay.
\end{abstract}

\section{\underline{Introduction.}}

Many extensions of the Standard Model (SM) predict the existence of an extra $U^{\prime}(1)$ gauge symmetry group and its associated $Z^{\prime}$ boson which has been object of extensive phenomenological studies~\cite{langacker-rmp}. This  boson can induce flavor changing neutral currents (FCNC) at tree level through $Z^{\prime}q_{i}q_{j}$ couplings where $q_{i}$ and $q_{j}$ are up or down-type quarks. The flavor-violating parameters must fulfill  experimental constraints on FCNC~\cite{aranda1}. We focus on the $Z^\prime$ virtual effects we may analyze the impact of the FCNC through the single top quark production. We can use the mass difference  $\Delta M_D$ of the $D^{0}-\overline{D^{0}}$ mixing observed by the Babar  and Belle collaborations  to bound the strength of these couplings.

\section{\underline{The $Z^{\prime} tu_{i}$ couplings.}}
The FCNC Lagrangian contained in the $SU_C(3)\times SU_L(2)\times U_Y(1)\times U^{\prime}(1)$ group is given by
\begin{equation}\label{general}
\mathcal{L}_{NC}=-eJ_{EM}^\mu A_{\mu}-g_1J_1^\mu Z_{\mu,1}-g_2 J_{2}^{\mu} Z_{{\mu},2}.
\end{equation}
$J^\mu_1$ is the weak neutral current and $J_2^\mu$ represents the new weak neutral current given as
\begin{align}\label{fotW}
J_{2}^{\mu}=\sum_{i,j} \overline{\psi^\prime_i} \, \, \gamma^{\mu} (\epsilon_{L{ij}}^{\psi}\, P_L+\epsilon_{R{ij}}^{\psi}\, P_R)\, \psi^\prime_j .
\end{align}
Since the interaction between the bosons $Z_{1}$ and $Z_2$ is too weak to be considered, there is no mixing between them, consequently their mass eigenstates are $Z^0$ and $Z^\prime$ respectively. Let us consider the $\epsilon_{L,R{ij}}^{u}$ matrix for the sector of quarks type up. Some models  assume this matrix as flavor diagonal and  non-universal. The FCNC couplings in the mass eigenstates basis can be read off as
\begin{align}\label{fotW2}
\Omega_{{L}ij}=\,g_2\,(V_{L}\, \epsilon_{L}^{u} \, V_{L}^{\dagger})_{ij},\,\,\,
\Omega_{{R}ij}=\,g_2\,(V_{R}\, \epsilon_{R}^{u} \, V_{R}^{\dagger})_{ij}.
\end{align}

\section{\underline{Bounding the  $Z^{\prime} tc$ couplings from $
D^{0}-\overline{D^{0}}$.}}

The Lagrangian containing the relevant information is
\begin{align}
\label{diagramas}
\mathcal{L}_{NC}^{Z^\prime q_iq_j}=&-\big[\overline{u}\,
\gamma^{\mu} (\Omega_{{L}uc} \, P_L+\Omega_{{R}uc} \,
P_R)\, c+\overline{c}\, \gamma^{\mu} (\Omega_{{L}cu} \,
P_L+\Omega_{{R}cu} \, P_R) \, u
\nonumber\\
+&\overline{u}\,
\gamma^{\mu} (\Omega_{{L}ut} \, P_L+\Omega_{{R}ut} \,
P_R)\, t+\overline{t}\,
\gamma^{\mu} (\Omega_{{L}tu} \, P_L+\Omega_{{R}tu} \,
P_R) \, u
\nonumber\\
+&\overline{c}\, \gamma^{\mu} (\Omega_{{L}ct} \,
P_L+\Omega_{{R}ct} \, P_R) \, t+\overline{t}\, \gamma^{\mu} (\Omega_{{L}tc} \,
P_L+\Omega_{{R}tc} \, P_R) \, c\big]Z^\prime_\mu.
\end{align}

From the unitary property of the $V_{L,R}$ matrices
\begin{equation}
|\Omega_{uc}| \approx |\Omega_{ut}\Omega_{ct}|,
\label{omega-relation}
\end{equation}
provided that $\epsilon_{tt}\ll 1$.
For simplicity we assume $\Omega$'s as real and $\Omega_{{L,R}\,q_iq_j}=\Omega_{{L,R}\,q_jq_i}$ and $\Omega_{L\,q_iq_j}=\Omega_{R\,q_iq_j}\equiv \Omega_{q_iq_j}$.
The tree-level amplitude can be written as
\begin{equation}
{\cal M}_{\rm tree}=-\frac{i\Omega^{2}_{uc}}{m_{Z^\prime}^2}\,\,\overline{u}\gamma^{\alpha}c\,\,\overline{u}\gamma_{\alpha}c.
\label{tree-level-amp}
\end{equation}
$\cal{M}_{\rm tree}$ amplitude can be related to a four-quark effective vertex accounted by the effective Lagrangian:
\begin{equation}
{\cal L}^{\rm tree}_{eff}=-\frac{\Omega^{2}_{uc}}{4m^{2}_{Z^\prime}}\left(Q_1+2 Q_2+Q_6\right),
\label{lag-tree}
\end{equation}
where a $1/4$ factor has been introduced to compensate Wick contractions. The $Q_i$ are dimension-six effective operators.

Analogously, the one-loop level amplitude is given by:
\begin{figure}[ht]
\centering
\subfigure[]{\includegraphics[scale=0.45]{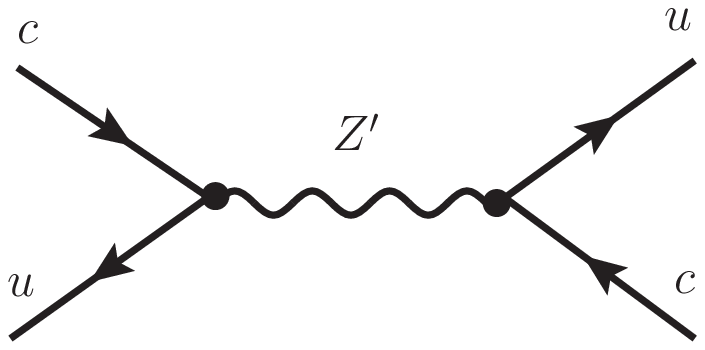}}\,\,\,\,\,\,
\subfigure[]{\includegraphics[scale=0.45]{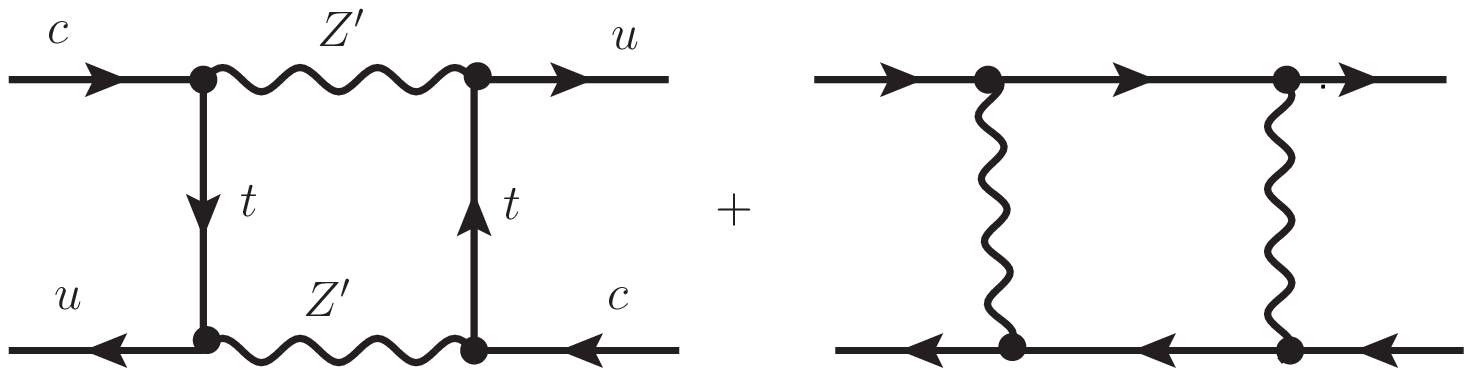}}
\caption{ (a) Tree diagram;(b) Box diagrams for $ D^{0}-\overline{D^{0}}$ mixing.}
\label{Figuras1}
\end{figure}
\begin{eqnarray}
\label{diagramas2} \mathcal{M}_{\rm box}=\,2\, \Omega^2_{tu}\, \Omega^2_{tc}\, \int \frac{d^{\,4}k}{(2\,\pi)^{4}}\,  \frac{[\overline{u}\, \gamma^{\lambda}\, (k^{\alpha} \gamma_{\alpha}+m_{t})\, \gamma^{\nu}\, c]\, [\overline{u}\, \gamma_{\nu}\, (k^{\alpha} \gamma_{\alpha}+m_{t})\, \gamma_{\lambda}\, c]}{(k^{2}-m_{t}^{2})^{2} (k^{2}-m_{Z^\prime}^{2})^{2}}.
\end{eqnarray}
After some algebra we arrive at the $\cal{M}_{\rm box}$ amplitude which can be related to a four-quark effective vertex accounted by the effective Lagrangian:
\begin{eqnarray}
\label{diagramas5} \mathcal{L}_{eff}^{\rm box}=-
 \frac{\Omega_{tu}^{2}\, \Omega_{tc}^{2}}{64 \pi^{2} m_{t}^{2}}\,  \big[f(x)\,
 (4Q_{1}+32 Q_{2}+4Q_{6})+g(x)\,(8 Q_{3}+4 Q_{4}+Q_{5}+4 Q_{7}+Q_{8})\big],
\end{eqnarray}
where a $1/4$ factor has been introduced to compensate Wick contractions; $f(x)$ and $g(x)$ are loop functions given as
\begin{eqnarray}
\label{diagramas4} f(x)=\frac{1}{2}\,
 \frac{1}{(1-x)^{3}}\,  [1-x^{2}+2\,x\,\log x],\, \, \, g(x)=\frac{2}{(1-x)^{3}}\,  [2(1-x)+(1+x)\,\log x].
\end{eqnarray}
with $x=m_{Z^\prime}^{2}/m_{t}^{2}$.
The mass difference $\Delta M_D$ provided by the $D^{0}-\overline{D^{0}}$ meson-mixing system is $\Delta M_{D}=\frac{1}{M_{D}}\,  Re \langle \overline{D^{0}}| \mathcal{H}_{eff}=-\mathcal{L}_{eff}|D^{0} \rangle.$
The effective Lagrangian is $\mathcal{L}_{eff}={\cal L}^{\rm tree}_{eff}+{\cal L}^{\rm box}_{eff}$ and $M_D$ is the $D^0$ meson mass.
Using the modified vacuum saturation approximation~\cite{pakvasa} we have:
\begin{eqnarray}
\Delta M_{D}=\frac{\Omega^{2}_{uc} f^{2}_{D}M_DB_D}{12 m^{2}_{Z^\prime}} \left[ 1+\frac{x}{8\pi^2}\big(32 f(x)-5g(x)\big)\right],
\label{deltam}
\end{eqnarray}
We used the relation in (\ref{omega-relation}), $B_D$ is the bag model parameter and $f_D$ represents the $D^0$ meson decay constant.
We can see from  Eqs.~(\ref{lag-tree}), (\ref{diagramas5}) and (\ref{deltam}) that the main contribution to $\Delta M_D$ comes from the tree-level amplitude while the contribution coming from the box amplitude  is of approximately  17\%-19\% in the range of $800\,\, \mathrm{GeV}\leq m_{Z^\prime}\leq 3000\,\, \mathrm{GeV}$.
Taking $B_D\sim1$, $f_D=222.6$ MeV and $M_D=1.8646$ GeV and considering that $\Delta M_D$  does not exceed the experimental uncertainty
\begin{equation}
|\Omega_{uc}|<\frac{3.6\times 10^{-7}m_{Z^\prime}{\rm GeV}^{-1}}{\sqrt{1+\frac{x}{8\pi^2}\left(32 f(x)-5g(x)\right)}},
\label{tree-contr}
\end{equation}
\begin{figure}[ht]
\centering
\includegraphics[scale=0.5]{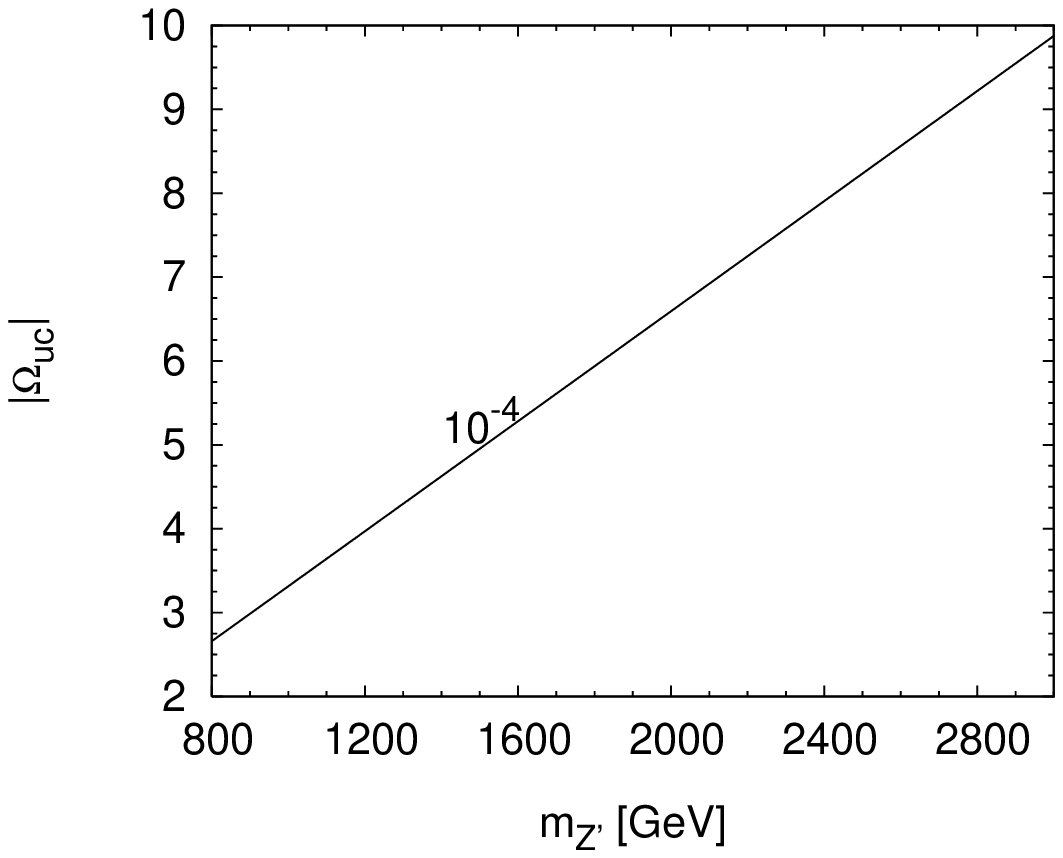}
\caption{Behavior of $|\Omega_{uc}|$ coupling  as a function of $Z^{\prime}$ boson mass.}
\label{Figuras2}
\end{figure}
Taking $m_{Z^\prime}=1$ TeV we obtain a bound $|\Omega_{tc}\Omega_{tu}|<3.31 \times 10^{-4}$, moreover, we assume that $\Omega_{tc}=10\,\Omega_{tu}$, as it occurs for the absolute values of $U_{ts}, U_{td}$ elements in the CKM matrix. We found that  $|\Omega_{tc}|<5.75\times 10^{-2}$  and $|\Omega_{tu}|<5.75\times 10^{-3}$, which are of the same order of magnitude approximately than those obtained in Ref.~\cite{arhrib}.

\section{\underline{The process $e^+e^-\to Z^\prime \to tc$ at ILC collider.}}

We only take  the average of the chiral charges; the different values for the  charges are: $Q_L^u=0.3456$, $Q_R^u=-0.1544$, $Q_L^d=-0.4228$, $Q_R^d=0.0772$, $Q_L^e=-0.2684$, $Q_R^e=0.2316$ and $Q_L^\nu=0.5$ for the Sequential $Z$ model; $Q_L^u=\frac{1}{\sqrt{24}}$, $Q_R^u=\frac{-1}{\sqrt{24}}$, $Q_L^d=\frac{1}{\sqrt{24}}$, $Q_R^d=\frac{-1}{\sqrt{24}}$, $Q_L^e=\frac{1}{\sqrt{24}}$, $Q_R^e=\frac{-1}{\sqrt{24}}$ and $Q_L^\nu=\frac{1}{\sqrt{24}}$ for $E_6$ model; $Q_L^u=0.2749$, $Q_R^u=-0.1793$, $Q_L^d=-0.1093$, $Q_R^d=-0.0635$, $Q_L^e=-0.0321$, $Q_R^e=0.0137$ and $Q_L^\nu=0.3521$ for Average model~\cite{pakvasa}.

The Breit-Wigner resonant cross section is $\sigma(e^+e^-\to Z^\prime \to tc)=\frac{12\,\pi\,m_{Z^\prime}^2}{s}\frac{\Gamma(Z^\prime\to e^+e^-)\,\Gamma(Z^\prime\to tc)}{(s-m_{Z^\prime}^2)^2+m_{Z^\prime}^2\Gamma_{Z^\prime}^2}.\label{cross}$
For the decay width $\Gamma(Z^\prime\to tc)$  we obtain $ \Gamma(Z^\prime\to tc)=\frac{\left(2 m_{Z^\prime}^4-m_{t}^4-m_{Z^\prime}^2 m_{t}^2\right) {\Omega^2_{tc}}}{12\,\pi\, m_{Z^\prime}^3 }.$
\begin{figure}[ht]
\centering
{\includegraphics[scale=0.45]{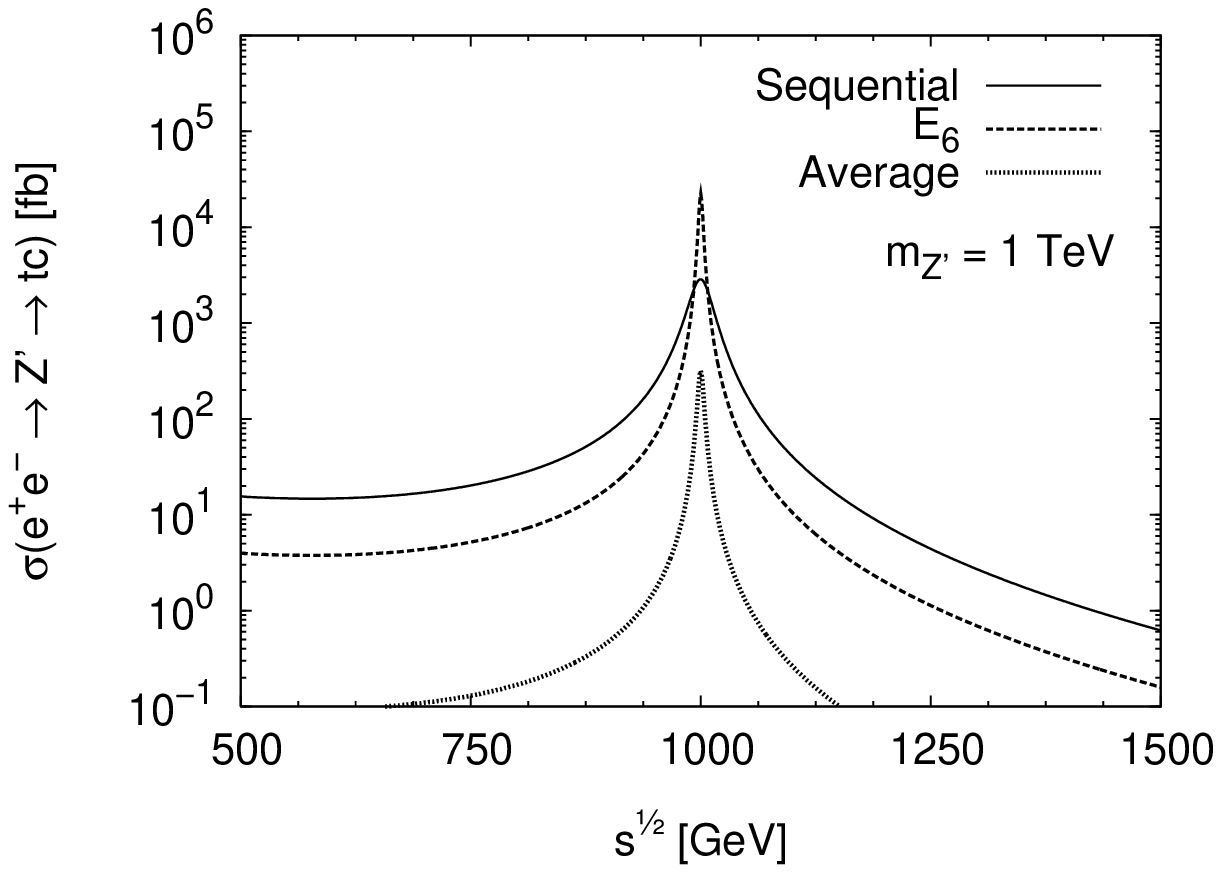}}\qquad
\caption{Cross section for $e^+e^-\to Z^\prime \to tc$ process as a function of $\sqrt{s}$ for $m_{Z^\prime}=1$ TeV.}
\label{cs}
\end{figure}

\begin{figure}
\centering
\includegraphics[scale=0.5]{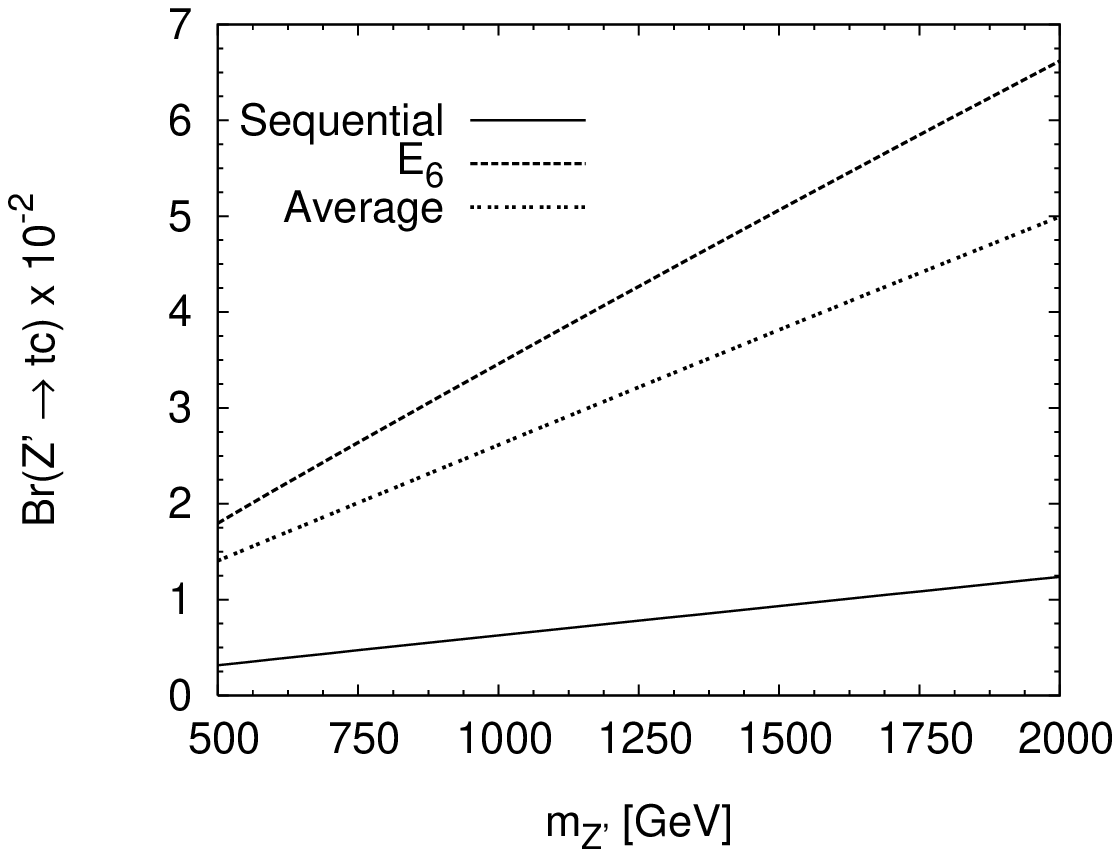}
\caption{\label{BR}The branching ratio for $Z^\prime \to tc$ decay.}
\end{figure}

We can predict around $10^7$ events just at the resonance for the $E_6$ model.
For the sequential $Z$ model it is expected to obtain around $10^6$ events.
For the average of the two models, it is expected around $10^5$ events.
We obtain that the associated branching ratio is of the order of $10^{-2}$. The production of around $10^4$ $t c$ events predicted in  Ref.~\cite{arhrib}, or  similar results in Ref.~\cite{cakir}, predicted for the Compact Linear Collider calculated at the resonance, can be compared with our predictions and find that ours  are bigger in  1  and   3 orders of magnitude for the average    and  the $E_6$ model, respectively. We found  that it will be produced around  $10^3$ $t c$ events   for a Higgs mass of the order of top quark mass, which is two orders of magnitude less than the average prediction, calculated at the resonance. In relation to the values we have found for  the branching ratios $Br(Z^\prime\to t c)\sim 10^{-2}$ and $Br(Z^\prime\to t u)\sim 10^{-4}$ calculated at the resonance, we can mention that these values are one order of magnitude less restrictive than corresponding  branching ratios obtained in the model 3-3-1~\cite{tosca}.

\section{\underline{Conclusions.}}
We have bounded the strength of the flavor-violating $Z^\prime tc$ coupling using the experimental results coming from the $D^{0}-\overline{D^{0}}$ meson mixing system. For a $m_{Z^\prime}=1$ TeV we found that $|\Omega_{tc}|<5.75\times 10^{-2}$. We have calculated the cross section for the $e^+e^-\to Z^\prime \to tc$ process in the ILC collider scenario; where we found an estimation around $10^7$ events for a luminosity of $500$ $\mathrm{fb}^{-1}$ in the context of $Z^\prime$ boson predicted by the $E_6$ model. According to our results the $tc$ flavor violation effect mediated by a $Z^\prime$ boson from the $E_6$ model is more favorable of being observed than that predicted in the sequential model one. This behavior is also repeated for the branching ratio of the $Z^\prime\to tc$ decay.
%
%

\protect\label{fig:thisdoc}

\end{document}